\documentstyle[12pt,prl,aps,psfig]{revtex}

\begin{document}
\draft
\title{Regular and Random Magnetic Resonance Force Microscopy Signal with a Cantilever Oscillating Parallel to a Sample Surface}
\author{G.P. Berman$^1$, V.N. Gorshkov$^{1,2}$, and V.I. Tsifrinovich$^3$}
\address{$^1$ Theoretical Division, Los Alamos National
Laboratory, 
Los Alamos, New Mexico 87545}
\address{$^2$ Department of Physics, Clarkson University, Potsdam, NY 13699}
\address{$^3$ IDS Department, Polytechnic University, Brooklyn, NY 11201}
\maketitle
\vspace{5mm}
\begin{abstract}
We study theoretically the magnetic resonance force microscopy (MRFM) in oscillating cantilever-driven adiabatic reversals (OSCAR) technique, for the case when the cantilever tip oscillates parallel to the surface of a sample. The main contribution to the MRFM signal is associated with a part of the resonance slice near the surface of the sample. The regular (approximately exponential) decay of the MRFM signal is followed by the non-dissipating random signal. The Fourier spectrum of the random signal has a characteristic peak which can be used for the identification of the signal.
\end{abstract}
\section{Introduction}
Inspired by the recent progress in magnetic resonance force microscopy (MRFM) based on oscillating cantilever-driven adiabatic reversals (OSCAR) technique \cite{1,2,3},
we consider here the theory of the OSCAR MRFM. In the OSCAR MRFM the oscillating cantilever with the ferromagnetic particle on its tip produces oscillating dipole field along the direction of the permanent external magnetic field. The rf magnetic field is applied in the plane perpendicular to the permanent external field. In the rotating system of coordinates the effective magnetic field changes its direction with the frequency of the oscillating cantilever. If the conditions of adiabatic reversal are satisfied for a spin of a sample then the spin will follow the direction of the effective magnetic field. The region of the sample where the effective magnetic field periodically reverses its direction is called the resonant slice. The spins of the resonant slice experience cyclic adiabatic reversals and produce the back effect on the cantilever: they cause the shift of the cantilever vibration period. In this paper we call this shift "the MRFM signal".
In previous works \cite{4,5,6} we have studied the ``perpendicular oscillations'': the cantilever tip oscillates perpendicular to the surface of a sample. In its advance to a single-spin detection the OSCAR MRFM technique relays on the ``parallel oscillations'', which allow one to reduce a spacing between the cantilever tip and the sample. Thus, the OSCAR MRFM theory should be extended to the parallel oscillations setup. This paper represents such an extension. We show that the thermal high frequency cantilever vibrations cause approximately exponential decay of the initial regular MRFM signal. However the MRFM signal does not disappear. It transforms into the random MRFM signal. The Fourier spectrum of the random signal has a characteristic peak, which can be used for the identification of the signal.
\section{Equations of motion}
We assume that a ferromagnetic spherical particle is attached to the cantilever tip and oscillates along the $x$-axis which is parallel to the surface of the sample. (See Fig. 1.) In the equilibrium position the center of the ferromagnetic particle is at the origin. The permanent external magnetic field $\vec {B}_{ext}$ points in the positive $z$-direction, and the {\it rf} magnetic field $\vec{B}_1$ rotates in the $x-y$-plane. 
The motion of the center of the ferromagnetic particle can be described by the equation
$$
\ddot x_c+x_c+\dot x_c/Q=f(\tau),\eqno(1)
$$
where $x_c$ is the dimensionless coordinate of the center (in terms of the oscillation amplitude of the cantilever, $x_0$), $Q$ is the effective quality factor of the cantilever, and we use the dimensionless time $\tau=\omega_ct$, where $\omega_c$ is the cantilever fundamental frequency. The dimensionless force $f(\tau)$ is produced by all magnetic moments of the ``resonant slice'' of a sample:
$$
f(\tau)=\sum_{k=1}^{\rm N}\eta_k\mu_{kz},\eqno(2)
$$
$$
\eta_k={{\mu_0}\over{4\pi}}{{3m\mu}\over{k_cx^5_0}}{{(\tilde x_k/\tilde r_k)(5{ z}_k^2/\tilde {r}_k^2-1)}\over{{\tilde r}^4_k}}.
$$
Here N is the number of magnetic moments in the resonant slice, ${\vec \mu}_k$ is the $k$-th magnetic moment (in terms of its magnitude $\mu$ which is the same for all magnetic moments), $k_c$ is the effective spring constant of the cantilever, $m$ is the magnetic moment of the ferromagnetic particle, $\tilde x_k=x_k-x_c$, $\tilde r_k=[(x_k-x_c)^2+y_k^2+z_k^2]^{1/2}$, $x_k$, $y_k$, $z_k$ are the coordinates of the $k$-th magnetic moment in terms of the oscillation amplitude $x_0$. We assume in (2) that only $z$-component of $\vec\mu_k$ influences the motion of the cantilever. 

\begin{figure}[t]
\centerline{\psfig{file=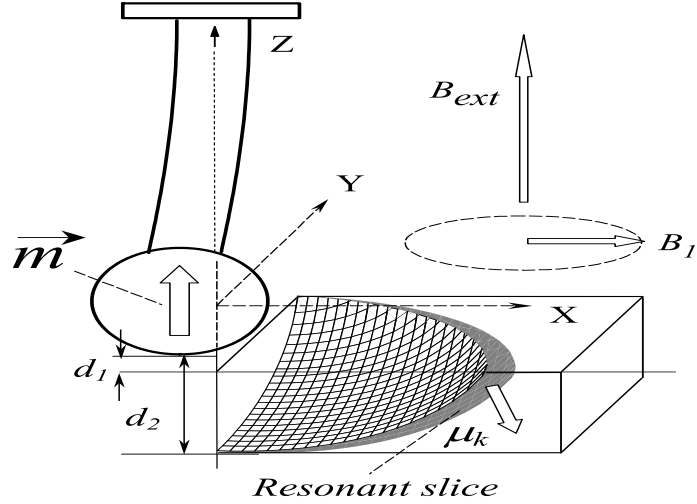,width=11cm,height=8cm,clip=}}
\vspace{4mm}
\caption{MRFM setup with the ferromagnetic particle oscillating parallel to the sample surface. $B_{ext}$ and $B_1$ are the permanent and rotating {\it rf} magnetic fields; $\vec m$ is the magnetic moment of the ferromagnetic particle;
$\vec\mu_k$ is the $k$-th magnetic moment in the resonant slice; $d_1$ is distance between the bottom of the spherical ferromagnetic particle to the sample surface; $d_2$ is distance between the bottom of the spherical ferromagnetic particle to the center of the resonant slice. The ``parallel setup'' was used in the experiments [2,3].}
\label{fig:1}
\end{figure}

The motion of the $k$-th magnetic moment of a sample in the rotating frame can be described by the equations:
$$
\dot \mu_{kx}=-\Delta_k \mu_{ky},\eqno(3)
$$
$$
\dot \mu_{ky}=\Delta_k \mu_{kx}-\varepsilon\mu_{kz},
$$
$$
\dot \mu_{kz}=\varepsilon\mu_{ky},
$$
$$
\Delta_k=(\gamma B_{ext}-\omega)/\omega_c+{{\mu_0}\over{4\pi}}{{\gamma m}\over{\omega_cx^3_0}}{{3 z^2_k/\tilde{r}^2_k-1}\over{{\tilde r}^3_k}}.
$$
Here $\gamma$ is the gyromagnetic ratio of the spins in the sample, $\omega$ is the frequency of the {\it rf} field, $\varepsilon=\gamma B_1/\omega_c$, and we neglect the transverse components of the dipole field produced by the ferromagnetic particle. 

The thermal vibrations of the cantilever tip with the frequencies close to the Rabi frequency $\gamma B_1$ cause the deviation of the magnetic moment from the direction of the effective magnetic field and the decay of the regular MRFM signal \cite{5,6,7,8,9}. To take into account this effect we make a substitution in the expression for $\Delta _k$ in (3):
$$
 x_c\rightarrow x_c+\delta x_c,\eqno(4)
$$
where $\delta x_c$ describes the thermal vibrations of the cantilever tip
$$
\delta x_c=\sum_n{{2a_n}\over{x_0}}\cos(\omega_n\tau+\Psi_n).\eqno(5)
$$
Here $a_n$ is the amplitude of the thermal vibrations for a cantilever mode $n$, $\omega_n$ is the eigenfrequency of the mode $n$ (in the units of $\omega_c$), $\Psi_n$ is its phase. In our simple model we assume that the amplitude $a_n$ can be found from the equipartition theorem
$$
m_c\omega^2_na^2_n=2k_BT,\eqno(6)
$$
where $m_c$ is the cantilever mass, and the phase $\Psi_n$ is the random function of time. The factor 2 in (5) appears as the amplitude of the tip vibrations for a uniform cantilever is twice the amplitude of the mode \cite{10}. (For a non-uniform cantilever this factor depends on $n$ \cite{11}.) We ignore the influence of the thermal vibrations on the value $\eta_k$ in (2).
\section{Numerical simulations}
In our numerical simulations we used the following parameters: the fundamental cantilever frequency $\omega_c/2\pi=7$kHz, the effective spring constant $k_c=10^{-4}$N/m, the effective quality factor $Q=5\times 10^4$, the amplitude of the cantilever vibrations $x_0=10$nm (to model the action of the feedback technique in the OSCAR MRFM, our computer algorithm increased the value $x_c$ to ``1'' every time when the cantilever tip passed the maximum value of $x_c$), the radius of the ferromagnetic spherical particle $R=200$nm, the magnetic moment of the sphere $m=2.5\times 10^{-14}$J/T, the distances from the bottom of the sphere to the sample surface and to the center of the resonant slice $d_1=220$nm and $d_2=300$nm (see Fig. 1), the amplitude and the frequency of the {\it rf} field $B_1=3\times 10^{-4}$T and $\omega/2\pi=3$GHz, the magnetization of the sample is $0.9$A/m. The resonant slice boundaries were found from the condition $\Delta_k(x_c=\pm 1)=0$. 

Note that for two spins with coordinates ($x,y,z$) and ($-x,y,z$) change of the $z$-component of the dipole field caused by the cantilever displacement has an opposite sign. Let, for example, the $z$-component of the effective field for these two spins is zero when the ferromagnetic particle is at the origin ($x_c = 0$). If $x_c\not=0$ the effective magnetic field on the first spin is opposite to that on the second spin. 

\begin{figure}[t]
\centerline{\psfig{file=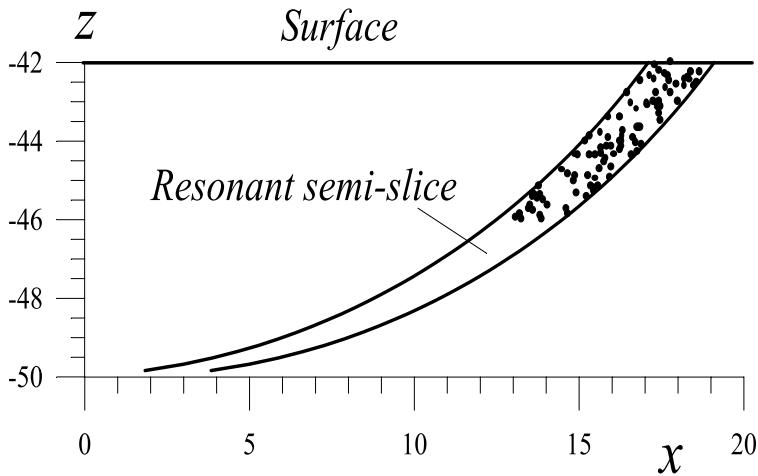,width=11cm,height=8cm,clip=}}
\vspace{4mm}
\caption{The cross-sectional area of the resonant semi-slice in the $x-z$-plane.  Dots show random distribution of magnetic moments.}
\label{fig:2}
\end{figure}
If the initial direction of the two spins relative to the external magnetic field is the same, these spins will have an opposite direction relative to the local effective field. Thus, the two spins induce the MRFM signal of the opposite sign. If the spins are uniformly distributed in the resonant slice, the net MRFM signal disappears. That is why in our simulations we assume that spins occupy only the resonant semi-slice $x > 0$.
The cross-sectional area of the semi-resonant slice in the $x-z$-plane is shown in Fig. 2. 

\begin{figure}[t]
\centerline{\psfig{file=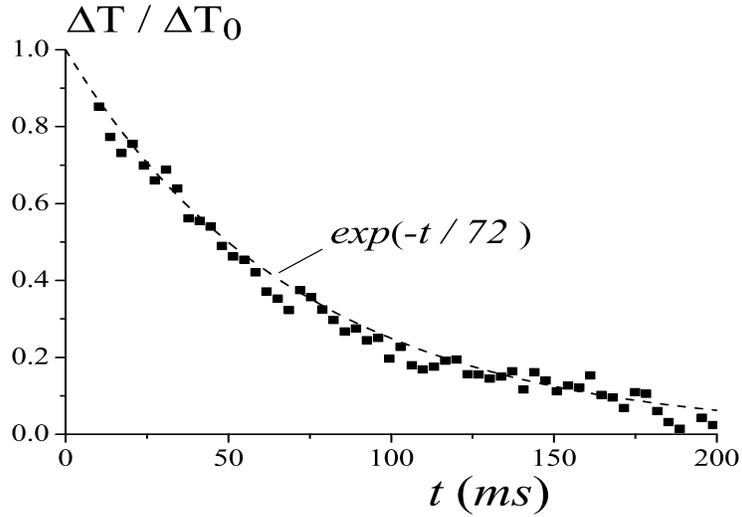,width=11cm,height=8cm,clip=}}
\vspace{4mm}
\caption{The decay of the regular MRFM signal. The temperature is 20K. The number of magnetic moments in the resonant semi-slice N = 100.
}
\label{fig:3}
\end{figure}

\begin{figure}[t]
\centerline{\psfig{file=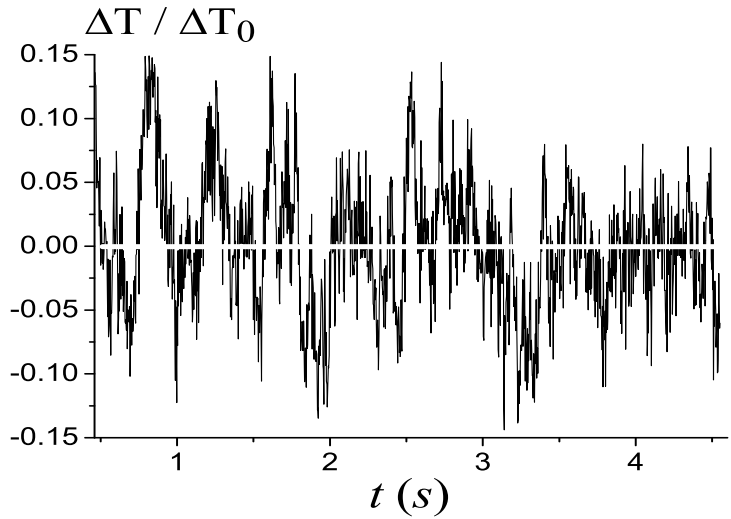,width=11cm,height=8cm,clip=}}
\vspace{4mm}
\caption{The random MRFM signal which follows the regular signal shown in Fig. 3.}
\label{fig:4}
\end{figure}

The magnetic moments in our model are randomly distributed inside the semi-resonant slice. We assume that initially all magnetic moments point in the positive $z$-direction, and the cantilever tip is in its right end point, $x_c=1$.

In our computer simulations the phases $\Psi_n$ in (5) were changed randomly between $0$ and $2\pi$, with random time intervals between two successive ``jumps''. The time interval between the phase jumps was taken randomly between $8.3T_R$ and $14T_R$, where $T_R=2\pi/\gamma B_1$ is the Rabi period. In Eq. (5), we took into consideration 25 cantilever modes in the vicinity of the Rabi frequency.

\begin{figure}[t]
\centerline{\psfig{file=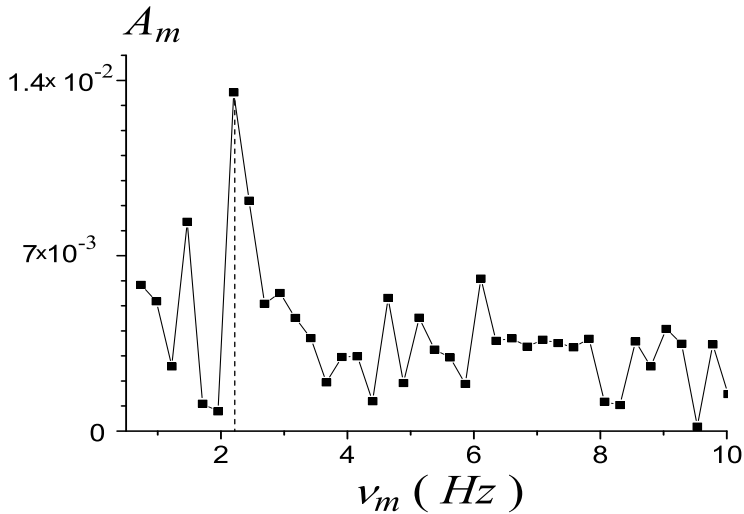,width=11cm,height=8cm,clip=}}
\vspace{4mm}
\caption{The Fourier spectrum of the random signal; $\Delta T(t)/\Delta T_0=\sum A_m\cos(\nu_mt+\Phi_m)$.}
\label{fig:5}
\end{figure}

Next, we describe the results of our simulations. The main contribution to the MRFM signal is associated with the magnetic moments near the surface of the sample. The regular MRFM signal decays approximately exponentially. The regular MRFM signal is followed by the non-dissipating random signal. The random signal has its characteristic signature: the major peak in its Fourier spectrum.

Figs 3-5 illustrate the results of our computer simulations. Fig. 3 shows the decay of the regular signal $\Delta T/\Delta T_0$, where $\Delta T$ is the shift of the cantilever period caused by the magnetic force, $\Delta T_0$ is the initial value of $\Delta T$. Fig. 4 demonstrates the random signal which follows the regular signal shown in Fig. 3. Finally, Fig. 5 shows the Fourier spectrum of the random signal with its characteristic peak.
\section*{Conclusion}
We have studied theoretically the OSCAR MRFM signal for the case when the cantilever tip oscillates along the axis parallel to the surface of the sample. In this case, the main contribution to the signal is associated with spins located in the part of the resonant slice near the surface of the sample. 
The main features of the OSCAR MRFM signal are the following: the regular (approximately exponential) decay of the MRFM signal is followed by the non-dissipating random signal.
 The Fourier spectrum of the random signal has a characteristic peak which can be used for identification of the signal.
\section*{Acknowledgments}
This work  was supported by the Department of Energy under the contract W-7405-ENG-36 and DOE Office of Basic Energy Sciences, and by the DARPA Program MOSAIC. 
{}
\end{document}